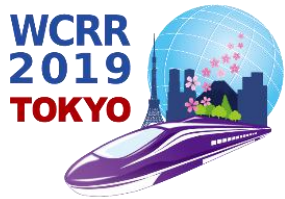

# A mechanical and economical based framework to help decision-makers for natural hazards and malicious events impact on infrastructure prevention.

Pierre-Jean TISSERAND[1,2], Muriel RAGUENEAU[2]

[1]*LMT, ENS Paris-Saclay, CNRS, Université Paris-Saclay*, Cachan, France

[2]*Division ISM, Département des OA, DG2I, SNCF RESEAU*, Saint-Denis, France

Corresponding Author: Pierre-Jean TISSERAND (pierre-jean.tisserand@reseau.sncf.fr)

## Abstract

Many studies in economics deal with the non-reliability cost to assess insurance fees or investment analyses, but none takes into consideration the mechanical aspect of reliability analysis. Other studies in mechanics give some tools and methods to carry out reliability analyses and fragility study. This study developed a framework where economical and mechanical considerations for infrastructure investment decision-making. The theoretical reasoning is here developed to couple mechanical reliability analyses, which are composed of fragility curves, and economical reliability analyses, which is based on resilience cost functions. This coupling is carried out with some probabilistic considerations, giving the concept of "probable cost of failure". The strength of this framework is that it can be used to analyze all possible critical components in a network with all possible natural hazards or malicious event or other undesired events which it is possible to assess its probability of occurrence. The results of the analysis are indicators of probable cost of failure of an infrastructure, which represents the insurance fee. These indicators can be computed for railway lines, for critical components, for events. This tool enables decision-makers to prioritize safety investments and to guide strategic choices. The next step of this study will be to develop smart data analysis tools, because of this framework needs and produces a lot of data, which must be smartly analyzed and presented.

1. ## Introduction

The first goal of the exposed methodology is to couple mechanical and economic studies on safety management to help decision-makers in infrastructure prevention face to natural hazards or malicious events. Relevant indicators are so needed to proceed. In mechanics, several indicators exist for safety assessment. A probability of failure can be computed through different models well-known by engineers. This indicator can be relevant for a system with a homogeneous non-reliability cost. When the cost of non-reliability is no longer homogeneous, this indicator and all those resulting from it are no longer relevant. Unthinking use can lead to absurdities: thus a bridge on an abandoned line will be equivalent to a high-speed line bridge. In economics, the observation still is the same. The economic indicators don't consider mechanical vulnerability and probability concept for probable cost of non-reliability consideration. The bad example could be to consider as critical a line with no risk, but with an important economic flow.[1]

The presented framework has the benefit to consider economic and mechanical results in one indicator, and those resulting from it lead to a classification of risks and critical components to prioritize investment.

## 2. General principle

Several kinds of coupling could be done between mechanics and economics on safety assessment, the one chosen is to rely on the probability of failure and the non-reliability cost (non-reliability = failure). The result is the probable cost of failure. The first equation (Eq.1) is the relationship between the probable cost of failure ($PCf$), the cost of failure Cf and the probability of failure ($Pf$).

$$PCf = Cf.Pf \quad \text{(Eq. 1)}$$

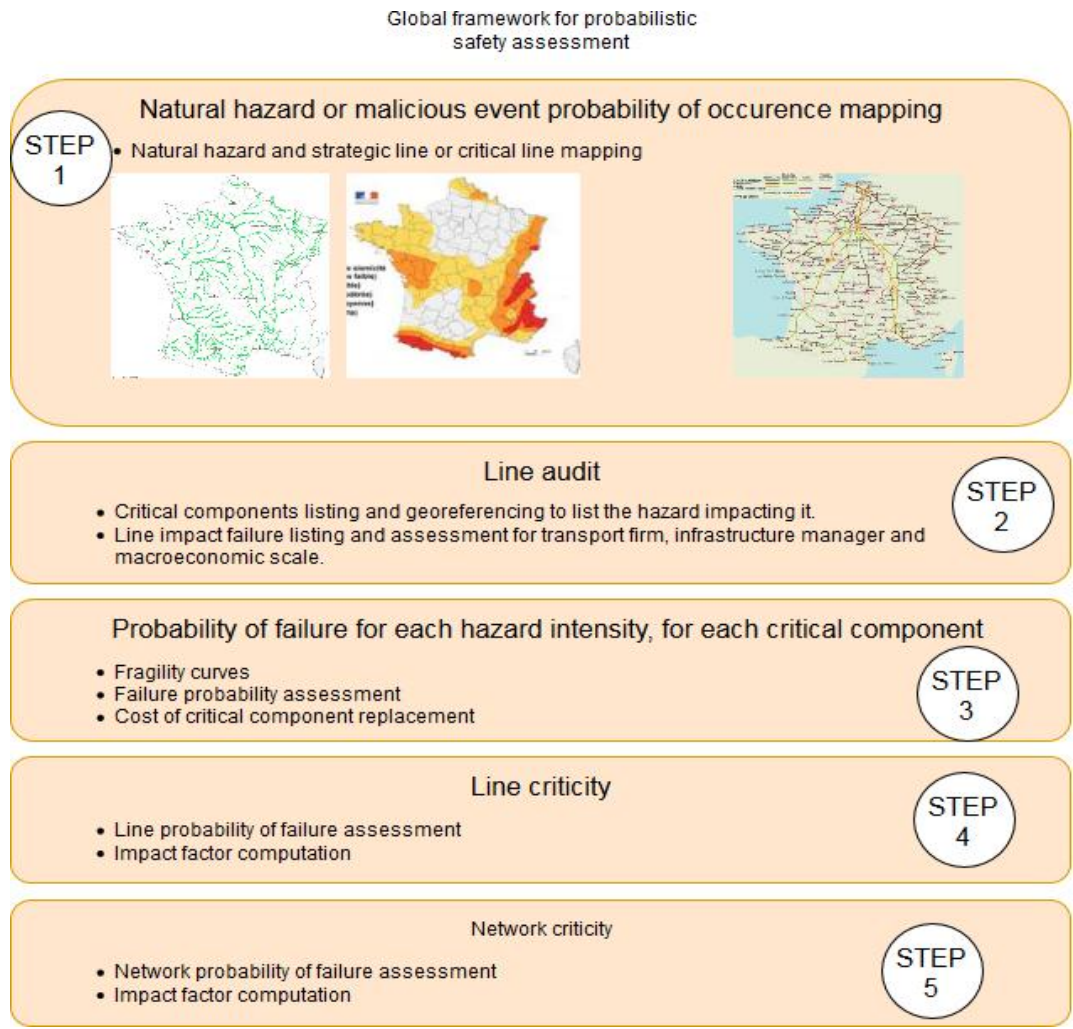

Fig. 1 : Methodology illustration

The figure 1 explains step by step the process of the probable cost of failure assessment.

## 3. Key points description

The main goal of the framework is to build indicators on the socioeconomic impacts of an event which implies the failure of a railway line taking into account mechanical considerations. In this study, the considered indicator is the probable cost of failure of a railway network. The probable cost of failure is assumed to be equal to equation (Eq.1).

In this part, the main theoretic considerations are explained.

In the first section, the notion of cost of failure will be introduced. In a second section, the probability of failure for a railway network will be developed, this need to introduce the third section, the probability of failure of a critical component, and the last section, the model of natural hazard or malicious events for mechanical computations.

## 3.1 Cost of failure [2], [4]

The cost of failure is the cost induced by the failure of an economic function. The cost of failure can be split into two parts: the direct cost of failure and the indirect cost of failure. The direct cost of failure is the cost of investment needed to solve the failure. In the case of a railway system, the direct cost can be the cost to repair or rebuild a bridge. The indirect cost is the cost induced by the impact of failure. In the case of railway system failure, it may be the cost of reimbursing tickets or the loss resulting from the service interruption. The indirect cost is strongly relied on the resilience mechanism. At the macroeconomic scale, the cost of failure should be measured through the GDP loss or increasing of time to reach locations.

The probable cost of failure proposed here is as insurance fee estimation. This is the probable cost of a failure for each year. For a series system, the probable cost is defined in figure 2.

Considering the Weiner diagram, the probable cost of failure of the series system is equation 2.

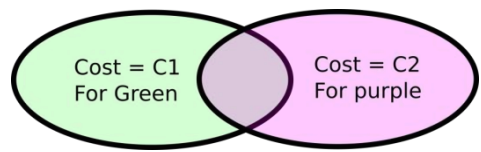

Fig. 2 : Weiner graph of cost probability

The direct cost of failure of a $n$ components series system :

$$PCf = \sum_{i=0}^{n} C_i . Pf_i \quad \text{(Eq. 2)}$$

The indirect cost can be compute directly in $C_i = Cd_i + Cind_i$, when $Cd_i$ is the direct cost of failure of critical component and $Cind_i$ the indirect cost of failure of critical component $i$, or $Pcf$ can be split in $Pcf = Pcfd_i + Pcfind_i$ with homogeneous indirect cost of failure. A first partial indicator can be compute as the part of probable cost due to a specific component or event.

$$Im_i = \frac{C_i . Pf_i}{Pcf} \quad \text{(Eq. 3)}$$

One of the most simple ways to assess large scale indirect cost of failure is to surrogate.

The indirect cost of failure is the integral of the cost function on time of failure.

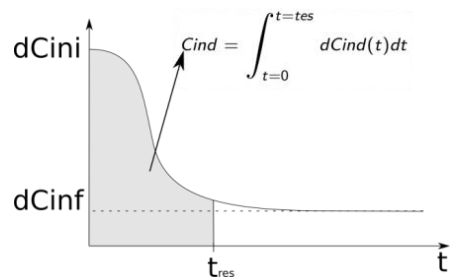

Fig. 3 : Indirect cost function

## 3.2 The probability of failure of a railway network.

The probability of failure of a network can be studied as the probability of failure nodes relied on lines. The railway system has a particularity, the lines are composed of a series of critical components. The critical component is a component which the failure implies the system failure.

In other words, the railway failure can be due to any of this critical component. When Pf is the probability of failure of the railway line, it is explained by the equation (Eq.4)

Considering $n$ critical on a line with $Pf_i$ the probability of failure of each component, the total probability for a series system is Eq.4.

$$Pf = 1 - \prod_i^n (1 - Pf_i) \quad \text{(Eq. 4)}$$

Two nodes can also be rely to 2 lines. This parallel configuration can be calculated with another equation with $Pf_i$ the probability of failure of the line $n$.

$$Pf = \prod_i^n Pf_i \quad \text{(Eq. 5)}$$

The next step is to compute the probability of failure of each critical component.

## 3.3 The probability of failure of a structure

The present study focuses on structure as bridge or tunnel. This work can be transposed on other components.

The probability of failure of a structure may be expressed through fragility curves. Fragility curves are a representation of conditional probability failure function of event intensity. Conditional probability of failure is defined by equation 6 when $FI$ is the failure indicator, $FL$ is the failure limit and $EI$ the event intensity.[3]

$$Fragility = P(FI \geq FL|EI) \quad \text{(Eq. 6)}$$

So for a giving $EI$, with a probability of occurrence of $EI$ $P(EI)$, the probability of failure is equation 7.

$$Pf_i = P(FI \geq FL|EI)P(EI) \quad \text{(Eq. 7)}$$

## 4. Example of results

In this part, an example of the previous methodology will be developed. For strategic reasons, the case is considered as fictitious. The following presents only results of the framework, in a step by step order.

The study will be a fictitious world, with plausible data, A-B valley. Here is a railway line cross country from A-town to B-town. The indicator the study will give to the decision-maker is the probable cost of failure of bridges. The line has 5 bridges, it crosses 2 rivers and is one two different seismic areas.

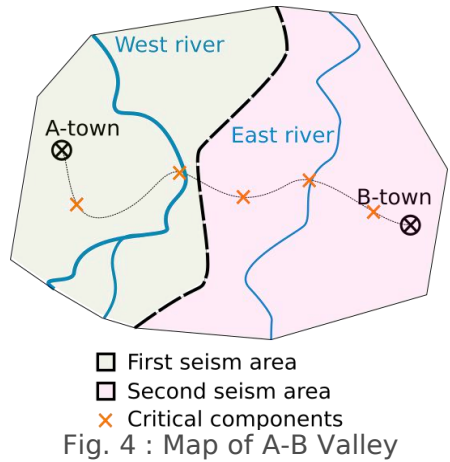
Fig. 4 : Map of A-B Valley

The seismology department gives the following data about the seismic areas.

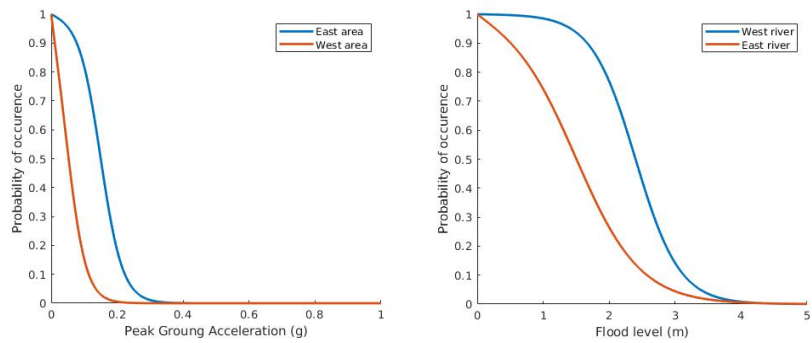
Fig. 5 : Probability of occurrence curves

The fragility curves of different bridges are here approached by log-normal surrogates.[3]

The study of this case gives the following curves for the probability of failure of each bridge.

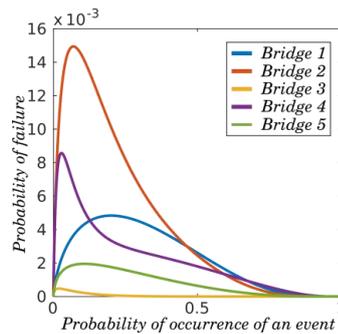
Fig. 6 : Probability of failure curves

Now, the decision-makers have to choose, if this is not imposed by standard, a back-period of natural hazards. The back-period here chosen is 1000 years.
An economic survey gives the following results for non-reliability cost for each bridge:

Table 1 : Cost of failure

| BRIDGE | Direct cost M€ | Indirect cost M€ |
|---|---|---|
| Bridge 1 | 1.56 | 6.13 |
| Bridge 2 | 1.24 | 2.46 |
| Bridge 3 | 1.72 | 5.13 |
| Bridge 4 | 1.05 | 9.75 |
| Bridge 5 | 1.35 | 3.52 |

Some indicators can be computed as the importance factor for each bridge and each event (Eq.3).

Table 2 : Importance factor for each bridges

|  | Bridge 1 | Bridge 2 | Bridge 3 | Bridge 4 | Bridge 5 |
|---|---|---|---|---|---|
| Total Importance factor | 0,28 | 0,27 | 0,01 | 0,38 | 0,06 |
| Importance factor only for seism event | 0.28 | 0.14 | 0.01 | 0.26 | 0.06 |
| Importance factor only for flood event | 0 | 0.12 | 0 | 0.13 | 0. |

Table 3 : Importance factor for each event

|  | Seism | Flood |
|---|---|---|
| Importance factor | 0.75 | 0.25 |

The probable cost for the line A-B is 0.06 M€. This indicator corresponds to naive insurance fee which should be saved each year. The results show that protection policy against floods should decrease the probable cost of failure of 25%. Many conclusions and decisions can be made from these indicators. The importance factor could also be computed line by line or part of a network by part of a network.

## 5. Conclusions and prospects

The framework meets the requirements and needs. It couples mechanical and economic considerations, gives indicators to sort and prioritize critical components, risks, and investment. The possibility to consider the shape of the system (series or parallel or combination of both) can extend the sorting by line or by part of the network.

The framework here presented is a piece of a larger work. To deliver a robust tool, many studies have to be carried out. At first, a great campaign of data gathering on events, critical components, mechanical and economical properties should be done. Then, the computation of fragility curves for all kind of events and critical components is a challenge which is already facing. And, the computation of indicators for large networks needs more than big data analysis tools, smart data analysis tools.